\documentclass[12pt]{article}
\usepackage{graphicx}
\usepackage{color}
\usepackage{amsmath}
\usepackage{amsfonts}
\usepackage{amssymb}
\usepackage{amscd}
\newcommand{\ourth}[2]{{\par\parindent=0pt\vspace{\baselineskip}{\bf
#1:}\ {\em #2}\vspace{\baselineskip}\par}\par}
\newcommand{\dA}{\fbox{\phantom{o}}}
\newcommand{\state}[2]{\begin{center}\vspace{\baselineskip}
\parbox{.85\textwidth}{{\bf #1:}\ {\em
#2}}\vspace{\baselineskip}\end{center}}
\begin{document}
\centerline{\bf The roots of scalar-tensor theory: an approximate history
\footnote{Contributions to the Cuba Workshop,
``Santa Clara 2004. I International Workshop on
Gravitation and Cosmology.''
}}\par
\begin{center}
{\bf Carl H. Brans}\par
Physics Department, Loyola University,New Orleans, LA\par
email: brans@loyno.edu
\end{center}\par
\begin{abstract}
Why are there no fundamental scalar fields actually observed in
physics today? Scalars are the simplest fields, but once we go
beyond Galilean-Newtonian physics they appear only in
speculations, as possible determinants of the
gravitational constants in the so-called  Scalar-Tensor theories
in non-quantum physics, and as Higgs particles, dilatons, etc.,
in quantum physics. Actually, scalar fields have had a long and
controversial life in gravity theories,  with a
history of deaths and resurrections. The first gravity
theory of scientific interest was developed by Newton,
using a scalar potential field. After developing special
relativity into which electromagnetism fit so nicely, it was
natural for Einstein and others to consider the possibility of
incorporating gravity into special relativity as a scalar
theory.  Of course, in its original form this effort was not
successful, but it did help in pointing the way to standard
Einstein general relativistic theory of gravity as a metric field.
However, the original investigation of a scalar field did
reinforce Einstein's interest in Mach's principle, suggesting an
influence of gravity on inertial mass.  Also, five dimensional
unified field theories as studied by Fierz, Jordan, and others
suggested  a spacetime scalar field that might well provide a
``varying'' gravitational constant. Even Einstein and Bergmann
were briefly interested in such possibilities. However, Dicke,
motivated by Dirac's numbers and Mach's principle, provided the
major driving force for theoretical and experimental
investigations of such a possibility.  While later
experimentation seems to indicate that if such a scalar exists
its influence on solar system size interactions is negligible,
other reincarnations of a scalar-tensor formulation have been
proposed in the contexts of dilatons in string theory and
inflatons in cosmology.  This paper presents a brief overview of
this history. A recent, and much more thorough, study of the
subject of scalar-tensor theories can be found in the book by
Fujii and Maeda, \cite{fm}. \end{abstract}
\section{Introduction} We begin by briefly reviewing the role of {\bf scalars} and
{\bf scalar fields} in physics.  Before
Einstein, the basic relativity principle in Galilean-Newtonian
physics required invariance in form of the laws of physics under
transformations of the Galilean group. Restricting ourselves to
Newtonian gravity and Maxwell's electromagnetism in this
context, we can easily find examples of scalars, such as mass,
electric charge, energy, etc.,  under the static (excluding
velocity transformations) affine subgroup of the full Galilean
group. However, when we allow constant velocity transformations,
the notion of scalars becomes a little less obvious. For example,
under the (constant) rotation group spatial intervals are clearly
scalars, but this is  not the case for non-trivial
velocity transformations, for which the spatial distance between
two events at the origin of one frame is zero as measured in that
frame, but not zero in another. Similar considerations apply to
speed, and thus kinetic energy. Clearly, the kinetic energy of a
particle is not a scalar under non-trivial velocity
transformations. Similarly, when we try to understand Maxwell's
electromagnetism in terms of a ``scalar'' and a ``vector''
potential, we find ourselves not able to formulate a consistent
theory invariant under constant velocity transformations, and
must rely on some fixed rest frame such as the ether. Of course,
these considerations are precisely those that led from Galilean
to Einsteinian special relativity, and a formulation of Maxwell's
electromagnetism in terms of a four-vector potential, with the
complete elimination of any scalar component of the
electromagnetic  potential. The next step, from special to general relativity
describes gravity in terms of a tensor, not a scalar, field.\par
Thus, while
scalars (constants) naturally abound, fundamental, i.e., not
derived,  scalar fields are only hypothetical to date. In
cosmology, pure Einstein theory uses only a 2-tensor, while in
quantum theory, the observed ``particles,'' such as quarks,
leptons, are represented by fermi spinors and the ``gauge forces''
are carried by boson vector fields. \begin{itemize}
\item{\em As of 2004, fundamental scalars appear only as
hypothetical, as yet unobserved, fields related possibly to the
gravitational or cosmological ``constants,'' dark energy,
inflatons, dilatons, or Higgs fields.} \end{itemize}
\par
In other words, nature seems to abhor using fields which have the
same value in all reference frames. This is surely a curious
fact.
\section{Special Relativistic gravity} In the early days of special
relativity, Einstein's first successful field theory was a
special relativistic re-formulation of Maxwell's electromagnetic
theory. Newtonian mechanics could be reformulated in terms
of force as a four-vector, ${\cal F}^\alpha,$ leading to a fully
Lorentz covariant theory of mechanics describing motion
parameterized by proper time, $x^\alpha(\tau)$
\begin{equation}
\frac{d}{d\tau}(m_i\frac{dx^\alpha}{d\tau})={\cal F}^\alpha.
\label{srfm}\end{equation}
where $m_i$ is a constant inertial mass, and of course,
\begin{equation}
\eta_{\alpha\beta}\frac{dx^\alpha}{d\tau}\frac{dx^\beta}{d\tau}=-1.\label{srfm2}\end{equation}
The constancy of $m_i$ and the consistency of (\ref{srfm}) and
(\ref{srfm2}) then require that the four-force be four-orthogonal
to the velocity,
\begin{equation}
\eta_{\alpha\beta}\frac{dx^\alpha}{d\tau}{\cal
F}^\beta=0.\label{srfm3}\end{equation}
This is clearly satisfied identically for the electromagnetic
four-force, ${\cal
F}^\alpha=F^{\alpha\beta}\eta_{\beta\gamma}\frac{dx^\gamma}{d\tau}$\par
 But what of a gravitational field theory? \begin{itemize}
\item{\em How does gravity fit into special
relativity?}\end{itemize} First, recall Newton's formulation in
terms of a Galilean scalar gravitational potential field:
\begin{equation}
\nabla^2\phi={\kappa\over 2}\rho_{ag},\label{ng1}\end{equation}
where $\rho_{ag}$ is mass density, with the $ag$ subscript
indicating that here the mass is {\bf a}cting as a source for
{\bf g}ravity. Also, $\kappa\equiv 8\pi G,$ and $G$ the usual
Newton constant. Using Galilean three-vector
notation, \begin{equation}
{\bf E}_g=-\nabla\phi,\label{ng2}\end{equation}
the equations of motion become
\begin{equation}
{d\over dt}\Big(m_{i}{d{\bf r}\over dt}\Big)=m_{pg}{\bf
E}_g.\label{ng3}\end{equation}
Here the $i$ subscript indicates {\bf i}nertial mass, while $pg$
corresponds to {\bf p}assive {\bf g}ravitational mass. Of course,
it was and is common to simply assume
\begin{equation}
{m_{ga}\over m_{gp}}=1,\label{defm2}\end{equation}
and
\begin{equation}
{m_{gp}\over m_{i}}=1.\label{defm4}\end{equation}
It is fairly easy to give an argument that momentum conservation
requires the satisfaction of (\ref{defm2}). On the other hand,
(\ref{defm4}) is less trivial, and corresponds to the
operationally significant
\begin{itemize}
\item{\bf Weak principle of
equivalence:}{\em Gravitational acceleration at a given point is
independent of  mass.}
\end{itemize}
So, as was common around 1900, let us temporarily assume
\begin{equation}
m_{ag}=m_{pg}=m_i=m=constant.\label{allmasses}\end{equation}
Finally, before leaving pre-Einsteinian gravity, we
   note that this potential, $\phi,$ has units
of velocity squared, so that in the standard relativistic choice
used in this paper, $c=1$, $\phi$ is dimensionless.\par
So, how do Einstein and his colleagues attack the problem of
integrating Newtonian gravity into special relativity?
Fortunately there are excellent, easily
readable, accounts of this process, \cite{jn1}, \cite{ray}.
What
might seem to be the most natural way to proceed? Simply assume
that gravity in special relativity will be described by a
4-scalar, $\phi,$ satisfying
\begin{equation}
\dA^2\phi={\kappa\over 2}\rho,\label{sg1}\end{equation}
\begin{equation}
{\cal F}_g^\alpha=-m\phi^{,\alpha},\label{sg2}\end{equation}
 (\ref{srfm}) as equation of motion. However,  (\ref{srfm3}),
applied to (\ref{sg2}) results in
\begin{equation}
{dx^\alpha\over d\tau}{\partial\phi\over\partial x^\alpha}=
{d\phi\over d\tau}=0.\label{sg3}\end{equation}
In other words, if we use (\ref{sg2}) and (\ref{srfm3})  the potential must
constant along the path of every particle, so the gravitational
force must necessarily be zero on every particle! Clearly, something is wrong here.\par
Consider the problem from the viewpoint of an action. A point
particle with path $z^\mu(\tau),$ and density,
$\delta^4(x^\nu-z^\nu(\tau).$ Here $\tau$ is proper time, so
\begin{equation}
 \eta_{\mu\nu}{dx^\mu\over d\tau}{dx^\nu\over
d\tau}=-1.\label{nm2}\end{equation}
As a guide, look at the electromagnetic equations, field and
particle motion, as derived from particle, field, and
interaction parts,
\begin{equation}\begin{array}{ll}
A_p+A_{em}+A_I & = -\int \Big(\int m\sqrt{-\dot z^\mu\dot z_\mu}
\delta^4(x^\mu-z^\mu(\tau))  d\tau\Big) d^4x \\ & {-1\over
16\pi}\int(A_{\mu,\nu}-A_{\nu,\mu})
(A^{\mu,\nu}-A^{\nu,\mu})d^4x+\\ & q\int\Big(\int\dot
z^\mu(\tau)A_\mu \delta^4(x^\nu-z^\nu(\tau))
d\tau\Big)d^4x.\end{array}\label{sg6}\end{equation}
Now consider a scalar gravitational modification of such a
formalism,
\begin{equation}\begin{array}{ll}
A_p+A_\phi+A_I & =-\int \Big(\int m\sqrt{-\dot z^\mu\dot z_\mu}
\delta^4(x^\mu-z^\mu(\tau))  d\tau\Big) d^4x \\
&
-{1\over \kappa}\int\phi_{,\mu}\phi^{,\mu}
d^4x\\
& -
\int \phi \Big(\int m\sqrt{-\dot z^\mu\dot z_\mu}
\delta^4(x^\mu-z^\mu(\tau))  d\tau\Big) d^4x
.\end{array}\label{sg7}\end{equation}
The field variation of this action  results in (\ref{sg1}) with
$\rho(x^\mu)=m\int\delta^4(x^\mu-z^\mu(\tau))d\tau$ in the
conformal gauge, (\ref{nm2}).    However, the variation over
the particle's variables, $z^\mu(\tau),\dot z^\mu(\tau)$ results
in something quite new,
namely,
\begin{equation}
{d\over d\tau}\big(m(1+\phi)\dot z^\mu(\tau)\Big)=-m\phi^{,\mu}.
\label{sg8}\end{equation}
 When the
four-vector equations of motion (\ref{sg8})
are expressed in terms of local coordinate time, it is clear that
local coordinate acceleration of a particle will depend not only
on the the particle's  kinetic energy, but also on a modified
inertial mass, $m(1+\phi)$,  thus violating the equal
acceleration principle, WEP.    Neglecting the
$\frac{d}{d\tau}\phi$ term, the usual expansion of the
left side of (\ref{sg8}) into local coordinate expressions gives
\begin{equation}
{d^2{\bf r}\over dt^2}\sim
-(1-v^2)\nabla\phi.\label{ca1}\end{equation}
Thus, the gravitational acceleration would depend on the
velocity, so spinning bodies would have smaller
accelerations in a gravitational field than non-spinning
identical ones, hot bodies than cold, etc.
Of course, this effect was too small to be noticed by early 20th
century technology, but naturally Einstein was disturbed by the
dependence of gravitational acceleration on internal structure
of the falling body occurring in this initial attempt to
``relativize'' gravity.
  \par
In a parallel vein,
  von Laue was looking into the models of internal stress in
extended bodies and found what we
now call the four dimensional stress-energy tensor, with
$T^{00}$ identified with energy density, and $T^{ij}=p^{ij}$ the
components of the spatial stresses on the body.
However, the application of a Lorentz velocity transformation to
such a tensor would  mix the purely spatial stress components into the
energy density, so that the energy density of a moving body
would depend on its internal
stress.  Thus, these internal stress components should contribute to the gravitational
mass.\par
 It might have
been something along these lines that motivated Einstein in 1907
to discount the appropriateness of a scalar special relativistic
theory of gravitation  because it did not allow ``...the inert
mass of a body to depend on the gravitational
potential.''\cite{e1} A related critique was formulated by
Abraham, \cite{jn1}.\par
Actually, Nordstr\"om, \cite{gn} suggested that  the
inertial mass might depend on $\phi$,
\begin{equation}
m=m_0\exp\phi,\label{gn1}\end{equation}
or for a weak field
\begin{equation}
m=m_0(1+\phi).\label{nord1}\end{equation}
   In fact, (\ref{sg8}) is related to Nordstr\"om's
(\ref{gn1}), with $\exp(\phi)\approx 1+\phi,$ to first order in
$\phi.$ Most importantly, the resulting equation of motion,
(\ref{sg8}), is consistent with (\ref{nm2}), as well as the
suggested field
and force equations, (\ref{sg1}) and (\ref{sg2}). \par
On the other hand, why associate $\phi$ with the mass? Why not
associate it with the metric, replacing
\begin{equation}
d\tau^2=(dt^2-dx^2-dy^2-dz^2),\label{scm1}\end{equation}
with
\begin{equation}
d\tau^2=\exp(2\phi)(dt^2-dx^2-dy^2-dz^2).\label{scm2}\end{equation}
This is the direction taken by Einstein
leading to his full general relativistic field equations using a
2-tensor, the metric as the potential.  In the spirit of
this paper this early scalar form for metric gravity, with
the scalar appearing as a metric ``dilation'' was very notable.
\par Historically, however, after its
brief, but passing, appearance in a Nordstr\"om model,
(\ref{scm2}), there seemed
 to be no room in physics for a scalar
field. But Nordstr\"om's suggestion (\ref{gn1}), led Einstein to
further pursue a {\bf Mach's Principle} in the sense of having
inertial mass depend on the gravitational interaction of all of
the other masses in the universe. We will briefly return to this
later.
\par
Of course, in parallel to relativity theory, quantum theory was
being developed, and scalar fields again appear in the context of
the Klein-Gordon equation. In turn,  this equation, and
its corresponding scalar field were replaced by Dirac equations.
As we now understand observed quantum physics, elementary
particles are fermions, satisfying Dirac equations, while forces
correspond to gauge fields which, while bosons, are spacetime
vectors rather than scalars. When we go beyond present day
observation, however, scalar fields may indeed return, as Higgs
bosons, dilatons, etc. We will mention these later.\par
As of the beginning of the 21st century, fundamental
scalar fields exist only as hypothetical structures in physics,
such as outlined in the following:
\begin{itemize}
\item {\bf Hypothetical non-quantum scalar fields }
\begin{itemize}
\item {\bf Scalar-tensor fields,} such as the JBD determinant of
$G$,
\item{\bf Inflatons}, scalar field to give rise to observed
anomalies of cosmological  expansion,
\end{itemize}
\item{\bf Hypothetical quantum scalar fields}
\begin{itemize}
\item {\bf Higgs particle}, quantum scalar field providing mass
by interactions with massless particles.
\item{\bf Dilatons, etc.,} quantum fields appearing in
superstring and M theory.
\end{itemize}
\end{itemize}
\section{The First Searches for Unified Field Theories}
The hunt for a theory unifying gravity and electromagnetism began
in the very earliest days of Einstein's general relativity. For
our purpose, the most significant was the 5-dimensional versions
associated with the names of Kaluza and Klein. Applequist et al
\cite{ACF} have prepared a convenient review and reprints of
original papers on the subject.\par
Briefly, KK theories enlarge the dimension of spacetime by one,
so that the metric has a form
\begin{equation} \gamma_{AB}=\begin{pmatrix} V^2 & V^2
A_\beta \\ V^2 A_\alpha & g_{\alpha\beta}+V^2 A_\alpha
A_{\beta}\\ \end{pmatrix}. \label{kk1}\end{equation}
 By restricting the five
dimensional transformation group appropriately,
$A_\alpha$ appear as the components of a spacetime
4-vector, with $V$ a spacetime scalar\footnote{Jordan et al,
\cite{SuW}, were able to characterize these transformations as
those of a projective group.}.  Furthermore, these
transformations could also account for electromagnetic gauge
transformations. Formally, this unification of spacetime and
gauge transformations made this sort of formalism highly
attractive, although the unobserved extra dimension was
generally regarded as an  obstacle to serious
consideration of the model. Also, there was the question of the
appearance of the unwanted scalar field, which Kaluza in 1921
\cite{k1}
described as ``noch ungedeutet.'' \par
But what are the field equations? By apparently natural extension
of the four dimensional Einstein equations, consider
\begin{equation}
\delta\int d^5x{\cal R}\sqrt{\vert
g^{(5)}\vert}=0.\label{kk31}\end{equation} These lead to spacetime
four dimensional equations as well as 4 spacetime-fifth
dimension equations and a single 5-5 equation, involving only
derivatives of $V=g_{55}.$ The spacetime equations are
\begin{equation}
R_{\alpha\beta}-\frac{1}{2}g_{\alpha\beta}R={V^2\over
2}(F_{\alpha\mu}F^{\mu}_{\ \ \beta}+{\eta_{\alpha\beta}\over
4}F_{\mu\nu}F^{\mu\nu})-({V_{,\alpha;\beta}\over
V}-{\eta_{\alpha\beta}\dA^2 V\over V}),\label{kk20s}\end{equation}
with $F_{\alpha\beta}$ the  electromagnetic components
derived from the potentials $A_\alpha$ as usual. These are the
standard Einstein equations with electromagnetic stress tensor
source, if we identify $V^2$ with 4 times the usual Newtonian
gravitational constant, $G$. However, in these equations $V$ may
not be constant and its derivatives also contribute.
\begin{itemize} \item{\em Equations (\ref{kk20s}) are the first
hint of a {\bf varying gravitational constant}.} \end{itemize}
\par
\section{Dirac's numbers}
Meanwhile, Dirac \cite{LNH}, building on the work of Eddington
and Milne, became intrigued by apparently ``coincidental''
approximate equality between important physical quantities
expressed in dimension free manner. Atomic scales can be deduced
from $\hbar, e$ and $m_p$, say the mass of the proton. Then an
atomic time(distance) scale is supplied by $T_a=R_a=e^2/m_a.$ On
the other hand we have the age(distance scale) and the mass of
the universe, $T_u=R_u, M_u$ as cosmological scales.  Finally,
we have the gravitational constant, $\kappa$. The resulting
dimensionless quantities could be approximately grouped into
powers of the incredibly large number, $10^{40}$,
\begin{equation}\begin{array}{rr}
\alpha = {e^2/\hbar}& \approx 10^{0},\\
{T_u/T_a} & \approx 10^{40},\\
{T_a/\kappa} & \approx 10^{40}, \\
{M_u/m_p} & \approx 10^{80}.\end{array}\end{equation}
For our purposes, the combination of these equations in the
following form is most important
\begin{equation}
\frac{1}{\kappa}\approx M_u/R_u.\label{Geq}\end{equation}
   \section{Scalar-Tensor Theories} \par
Perhaps the earliest work in this direction was pursued
independently by Jordan in Germany and Einstein and Bergmann in
the US beginning in the late 1930's. Of course, this work
proceeded under all of the terrible constraints of the second
world war  and the resulting isolation of the two groups.
Actually, Einstein and Bergmann apparently decided not to
proceed with the variable gravitational constant idea to the
point of publication. Bergmann \cite{pgbkk} reviews these
parallel efforts in his paper,  ``Unified field
theory with
fifteen field variables,'' from which we now quote:\par
\begin{quote}

In the spring of 1946, Professor W. Pauli turned over to the
author of this paper galleys of a paper by P. Jordan entitled
``Gravitationstheorie mit veranderlicher Gravitationszahl'',
which was to have appeared in the Physikalische Zeitschrift
sometime in 1945, but which was, of course, never published
because the Phys. Zeitschrift in the meantime ceased publication.
In this paper, Jordan attempted to generalize Kaluza's five
dimensional unified field theory by retaining $g_{55}$ as a fifteenth field
variable. Professor Einstein and the present author had worked
on that same idea several years earlier, but had finally rejected
it and not published that abortive attempt. The fact that another
worker in this field has proposed the same idea, and independently,
is an indication of its inherent plausibility. Therefore, it
seemed worthwhile to review these attempts to
``vary the constant of gravitation'' and to discuss the
possibilities inherent in geometries of this kind.
\end{quote}
Thus, independently of Einstein and Bergmann in the USA,
      Pascual Jordan and his colleagues   in Germany
\cite{SuW}
began an extensive look at Kaluza-Klein theories
with special concern for the possibility that the new
five-dimensional metric component, a spacetime scalar, might
play the role of a varying gravitational ``constant,''
as suggested by Dirac's (\ref{Geq}).
Certainly the resulting four-dimensional form of the field
equations can interpreted this way.  However, Jordan  and his
colleagues went beyond the 5-dimensional origins of this scalar
and proposed purely four dimensional field equations involving a
scalar field related to Newton's constant.  Later Brans and Dicke
\cite{bd} independently arrived at similar point.
However, for Brans and Dicke, Mach's ideas on inertial induction,
that the total mass distribution in the universe should
determine local inertial properties, were of prime concern. In
fact, Sciama\cite{Sci} had earlier proposed a model theory of
inertial induction. \par
Sciama's work provided a theoretical
model in which inertial forces felt during
acceleration of a reference frame relative to the ``fixed
stars,'' are  of gravitational origin.
From this assumption, Dicke argued that Mach's principle would
manifest itself in having the ratio of inertial to gravitational
mass depend on the average distribution of mass in the
universe. That is,
\ourth{Dicke's form of Mach's Principle}{The gravitational
constant, $\kappa,$ should be  a function of the mass
distribution in the universe.} Because of Dirac's large number
hypothesis in the form
\begin{equation} 1/\kappa\approx
M/R,\label{lnh1}\end{equation}
it seems that the reciprocal of the gravitational constant will
likely be the field quantity.  In other words, $1/\kappa$ itself
might be a field variable and satisfy a field equation with mass
as a source, something like
\begin{equation}
\mbox{`` $\dA$ ''}{1\over\kappa}=\rho.\label{st0}\end{equation}
So,  introduce
a scalar field, $\phi$, which will
play the role, at least locally and approximately,  of
the reciprocal Newtonian gravitational constant, $\kappa$.
\par
The usual Lagrangian for Einstein theory including matter
contains $\kappa$ directly multiplying the matter contributions.
Keeping the field directly coupled to matter would then
inevitably lead to changes in the local behavior of
matter, the local equations of motion, as a result of variations
in $\phi.$  So, in order to incorporate a Mach's
principle by way of a variable gravitational ``constant,'' we
need to look at further modifications in the form of the general
relativistic action. Begin with the
standard Einstein action as \begin{equation} \delta\int
d^4x\sqrt{-g}(R+\kappa L_m)=0,\label{sto-1}\end{equation} where  $L_m$ is the
``usual'' matter Lagrangian, a priori derived from some
particular classical or quantum model. Equation (\ref{sto-1}) is
clearly not enough since it provides no field equation for the
new field, $\kappa.$
\par
Before proceeding, we need to review some  aspects of the
famous (or infamous)  ``principle of equivalence.'' As usual we
neglect tidal forces, extended bodies, etc., in these idealized
models. Dicke often pointed out that we need to
distinguish two forms: \begin{itemize} \item{\bf WEP.} One form
asserts that all bodies at the same spacetime point in a given
gravitational field will undergo the same acceleration.  We will
refer to this as the ``weak'' equivalence principle, WEP. As it
stands, this does not exclude  possible effects of gravity
other than acceleration. \item {\bf SEP.} A stronger statement,
which actually is important to Einstein's general relativistic
theory of gravity, is that the {\it only} influence of gravity
is through the metric, and can thus (apart from tidal effects)
be locally, approximately transformed away, by going to an
appropriately accelerated reference frame.  This is the
``strong'' principle, SEP. \end{itemize}  An action of the form
in (\ref{sto-1}) with variable $\kappa,$ will clearly change the
geodesic equation for test particles, thus, possibly the WEP, and
even mass conservation.   As Dicke noted, the E\"otv\"os
experiment verifies  the WEP (but not the SEP), so, in the
1960's, we would like to at least modify (\ref{sto-1}) to agree
with the WEP. To ensure the geodesic equations for point
particles we  isolate $\kappa$ from matter in the
original (\ref{sto-1}) by dividing by it, \begin{equation}
\delta\int d^4x\sqrt{-g}(\phi R+ L_m)=0,\label{sto-2}\end{equation} where
we have replaced $\kappa\to 1/\phi.$    However, we should
note the following. While we seem to have  saved the geodesic
equations for test particles,
the motion of composite bodies is more complex. It turns out
that  the coupling of a new, universal field,
$\phi$ directly to the gravitational field gives rise to
potentially observable effects for the motion of matter
configurations to which gravitational energy contributes
significantly. This is now known as the``Dicke-Nordtvedt''
effect and has been investigated
in the earth-moon system with the lunar laser reflector, leading to possible
violations of even the WEP for extended masses.
These possibilities were not considered in the early
days. So, let us proceed to see what follows  from
(\ref{sto-1}). We need field equations for $\phi$ so
some action for this new field must
be supplied, \begin{equation} \delta\int d^4x\sqrt{-g}(\phi R+
L_m+L_\phi)=0.\label{sto-3}\end{equation}
We must note,
that by allowing a new, scalar, field, we are opening the door to
other consequences. Since gravity is
universally coupled to all physics, the direct
coupling of $\phi$ to geometry, $\phi R$,
means that $\phi$ is universally coupled in
some sense. \ourth{Consequences of (\ref{sto-3})}{ We are
allowing for a possible violation of the SEP, since gravity, the
universal interaction of mass, can influence local physics, not
only through geometry, but also by changing the local
universally coupled  $\phi$, thus changing internal
gravitational structure.}  The usual requirement that the field
equations be second order leads to \begin{equation}
L_\phi=L(\phi,\phi_{,\mu}).\label{sto-4}\end{equation} Apart from this, there
seem to be few {\it a priori} restrictions on $L_\phi.$  The standard choice for a scalar field, \begin{equation}
L_\phi=-\omega\phi_{,\mu}\phi_{,\nu}g^{\mu\nu},\label{sto-5}\end{equation}
results in a wave equation for $\phi$ with $R$ as source seems
natural.  However, the coupling constant $\omega$ would itself
then need to have the same dimensions as the gravitational
$\kappa$ that the new field is to replace! But, one of the
motivations for extending Einstein's theory is to eliminate the
dimensional constant, $\kappa.$  So we require that any new
coupling constant appear as dimensionless. An obvious natural
minimal choice is \begin{equation}
L_\phi=-\omega\phi_{,\mu}\phi_{,\nu}g^{\mu\nu}/\phi,
\label{sto-6}\end{equation}
in which the field $\phi$ has dimensions of inverse gravitational
constant,
\begin{equation}
[\phi]=[\kappa^{-1}].\label{sto-6.1}\end{equation}
In fact, we will soon see, (\ref{bd-6}), that this results in
field equations suggestive of (\ref{st0}).
 The form (\ref{sto-6})
leads to an action which is often referred to as the
``Jordan-Brans-Dicke,''  JBD, action,
\begin{equation}
\delta\int d^4x\sqrt{-g}(\phi R+ L_m
-{\omega\over\phi}\phi_{,\mu}\phi_{,\nu}g^{\mu\nu})=0.
\label{bd-1}\end{equation}
The variational principle, with standard topological and surface
term assumptions, results in
\begin{equation}
\delta_m\int dx^4\sqrt{-g} L_m=0,\label{bd-2}\end{equation}
\begin{equation}
\phi S_{\alpha\beta}= T_{(m)\alpha\beta}+\phi_{;\alpha;\beta}
-g_{\alpha\beta}\dA\phi+{\omega\over\phi}
(\phi_{,\alpha}\phi_{,\beta}-{1\over 2}g_{\alpha\beta}
\phi_{,\lambda}\phi^{,\lambda}),\label{bd-3}\end{equation}
\begin{equation}
\omega({2\dA\phi\over\phi}-{\phi_{,\lambda}\phi^{,\lambda}
\over\phi^2})=-R.\label{bd-4}\end{equation}
The first of these, (\ref{bd-2}), is the standard variational
principle for matter, leading to the same expression for matter
motion in terms of the metric. It thus (apparently) satisfies
the weak equivalence principle.  For example test particles,
(\ref{bd-2}), follow geodesics.  However, if the matter is
extended, not a point particle, this is may no longer be true,
even in standard general relativity.  However, for scalar-tensor
model, there is a  second order interaction of matter through the
scalar-metric coupling. This thus gives rise to  violations of
the weak equivalence principle.  In other words, extended bodies
of different mass may have different gravitational accelerations
at the same point in a gravitational field. Of course, we do
have the standard energy tensor for matter and resulting matter
conservations laws. This is a result of the choice because of
the free standing $L_m$ in (\ref{bd-1}), not directly coupled to
$\phi$, \begin{equation} T_{(m)\alpha\
;\beta}^{\phantom{(m)\alpha}\beta}=0. \label{bd-4.1}\end{equation}
We can couple $\phi$ directly to matter by
taking the trace of (\ref{bd-3}), solving for $R$.
The result is another form for (\ref{bd-4}),
\begin{equation}
\dA\phi={1\over (2\omega+3)}T_{(m)},\label{bd-6}\end{equation}
in which $T_{(m)}$ is the trace of the ordinary matter tensor.
It should be noted that traceless matter, such as null
electromagnetic fields, do not directly couple to $\phi.$ \par
Now, to look at the possible satisfaction of Dirac's
(\ref{lnh1}), consider a weak field model situation  with a
static spherical shell of mass $M$, radius $R$ and otherwise
empty universe this equation. The result
is \begin{equation} \phi\approx \phi_\infty+{1\over
4\pi(2\omega+3)}{M\over R}.\label{bd-6.2}\end{equation}
Dividing equation (\ref{bd-3}) by
$\phi$ results in an equation in which the ``ordinary'' matter
tensor, $T_{(m)\alpha\beta}$ is divided by $\phi$, which thus can
be identified with the local reciprocal gravitational constant.
Also, of course, the $\phi$ contributes its own stress energy
matter tensor to the right side of (\ref{bd-3}). If
$\phi_\infty$ is set zero as a default asymptotic condition,
then (\ref{bd-6.2}) is seen to be consistent with the Dirac
coincidence, (\ref{lnh1}).  A natural approximation to
(\ref{bd-6}) is to consider the effect of local matter over some
background $\phi_0$ equal to the present observed value,
\begin{equation}
\phi\approx \phi_0+\frac{1}{ 4\pi(2\omega+3)}\sum_{\mbox{local
m}}{m\over r}.\label{bd-6.3}\end{equation} This can be regarded as
an extension of Dirac's (\ref{lnh1}).
 \par In equation (\ref{bd-3}),
$T_{(m)\alpha\beta}$ are the components of the stress-energy
tensor for matter derived from the matter Lagrangian $L_m$ in the
standard fashion.  Grouping this term with the $\phi$ ones,
results in an interpretation of  \begin{equation}
S_{\alpha\beta}=(1/\phi)(T_{(m)\alpha\beta}+
T_{(\phi)\alpha\beta}),\label{bd-5}\end{equation}
as the {\bf total} source for the Einstein tensor in(\ref{bd-3}).
So,  $(1/\phi)$ does indeed act as a
generalized gravitational ``constant'', with both ordinary
matter and the field $\phi$ itself serving as sources for the
metric. Actually the the $\dA\phi$
term on the right hand side of (\ref{bd-3}), together with
(\ref{bd-6}) results in two occurrences of the matter tensor as
a source. Thus there could be some argument for renormalizing
$1/\phi$ as the  ``gravitational
constant'' multiplying ordinary matter as it contributes to the Einstein tensor. \par
Pascual Jordan and his
collaborators were the earliest serious investigators of
equations of this sort. Most of the work by Jordan and his group
is summarized in Jordan's book, \cite{SuW}. See also a more
recent review by Sch\"ucking \cite{SchPT}.  In addition to
surveying the projective UFT's motivation, Jordan's book contains
thorough studies of the  static, spherically symmetric
generalizations (the Heckmann solutions) of the Schwarzschild
solution as well as
cosmological solutions and other topics. \par
Equation (\ref{bd-1})  the brings to mind actions obtained by
conformal changes of the metric. So, it is natural to look at the
action of the local ``conformal group'' on the representations of
the theory.  Replace the metric,
$g_{\mu\nu}\rightarrow \bar{g}_{\mu\nu}=\psi g_{\mu\nu}$. Discarding the
surface (topological) part, (\ref{bd-1}) becomes
\begin{equation}
\delta\int d^4x\sqrt{-\bar{g}}({\phi\over\psi}
\bar{R}+{3\phi\over
2}{\vert\bar{\nabla}\psi\vert^2\over\psi^3}
-3\bar{\nabla}\psi\cdot\bar{\nabla}\phi/\psi^2+ L_m/\psi^2
-{\omega\over\phi\psi}\vert\bar{\nabla}\phi\vert^2)=0.
\label{bd-1y}\end{equation}
If $\psi$ is chosen to be $\phi$,
(\ref{bd-1y}) becomes
\begin{equation}
\delta\int d^4x\sqrt{-\bar{g}}(\bar{R}-(\omega+{3\over
2})\vert\bar{\nabla\alpha}\vert^2+e^{-2\alpha}L_m(\bar{g}))=0,\label{cf-1}\end{equation}
where $\phi=e^\alpha$.  It is easy to see that this variational
principle is  just the Einstein one for a massless
scalar field(dimensionless), $\alpha$, but universally coupled
to all other matter through the $e^{-2\alpha}$ factor.
These conformal rescalings of the metric constitute the
``metric gauge group.'' Thus (\ref{cf-1}) is an expression of
the theory in the ``Einstein gauge,'' as opposed to the original
(\ref{bd-1}), the ``Jordan'' gauge.  But there is more to the
conformal scaling than merely the formal expression of the
equations.  Most significantly, the universal coupling of
$\alpha$ to {\em all} matter in (\ref{bd-1y}) or (\ref{cf-1})
means that, in this metric, test particles will not follow
geodesics, nor have conserved inertial mass, etc., in the
Einstein gauge. In other words, conservation laws derived from
the matter tensor depend on the construction of that tensor from
the function multiplying $\sqrt{-g}$ in the action,
(\ref{bd-1y}) or (\ref{cf-1}).    Choosing the Einstein metric
in  (\ref{cf-1}) as the ``physical'' metric leads to significant
and observable violations of mass conservation and the WEP.
\par
In the 1960's and 1970's, Bob Dicke was a leading influence
influence in the push to experimentally test general relativity
in Einstein's original form as well as alternatives such
scalar-tensor generalizations\cite{Dic:64}. In fact, the
explosion of interest in relativity and gravitational theories
and tests was prompted at least in part by the presence of
theoretically viable alternatives to standard Einstein theory,
and Dicke's energetic promotion of them.  Also  NASA was coming
of age and searching for space related experiments of
fundamental importance.  The important bridge between theory and
experiment in gravitational theories was developed by Thorne,
Nordtvedt, Will and others \cite{will}. Their work provided
rigorous underpinnings for the operational significance of
various theories, especially in solar system context. An
important tool is  the parameterized post Newtonian (PPN)
formalism which provides theoretical standard for expressing the
predictions of relativistic gravitational theories in terms which
can be directly related to experimental observations.\par
 From
(\ref{bd-3}), it appears that the equations of scalar-tensor
theory approach those of standard Einstein theory as $\phi$
approaches a constant.  From (\ref{bd-6.2}) this would seem to
occur in the limit of large $\omega$. Of course, this equation
is just an approximation to a solution of (\ref{bd-6}) for an
asymptotically empty universe, with $\phi\to 0$ as boundary
condition. Actually, these comments obscure the need for
rigorous analysis for the action (\ref{bd-1}) as
$\omega\to\infty.$ This is not surprising since the limiting
dependence of solutions of field equations on parameters in
these equations is in general a complex problem with all of the
subtleties associated with the topology of a space of functions.
However, it is true that \state{Approach to standard Einstein}
{In the realm of solar system experiments, the predictions of a
theory of the form (\ref{bd-1}) approach those of standard
Einstein theory as $\omega\to\infty.$} So, tests of such
theories are often expressed as providing lower limits for
$\omega.$ For more details, see  \cite{dn}.
\par
As the experimental data on solar system gravitational
measurements come in, new limits on the value of the parameter
$\omega$ have become so large as to make the predictions of this
theory essentially equivalent to those of standard Einstein
theory.  In other words, from solar system experimentation it
seems that scalar-tensor modifications of standard Einstein
theory would necessarily differ insignificantly from the
standard.
\par
Gravitational radiation provides another arena for experimental
studies of gravitation. In 1975 the Hulse-Taylor binary pulsar
decay data\cite{HulTay:75} showed that gravitational radiation
can provide another tool for testing gravitational theories.
More recently, part of the justification for the LIGO
gravitational radiation study is to provide further comparison
of standard Einstein to alternative theories\cite{Wil:99}.
\par
In spite of the apparently unpromising solar system experimental
results, it turns out that  universally coupled, thus
gravitational, scalar fields continue to play important
roles in contemporary physics. David Kaiser\cite{Kai:05} has
given a review of this topic, comparing JBD and Higgs fields, for example.
We will briefly consider some of these
possibilities in the following sections.\par
\section{Dilatons}\par
As discussed in the introduction, it is surprising that scalar
fields do not seem to occur naturally in special relativistic,
pre-quantum physics. However, from the earliest days of quantum
theory, scalar quantum fields were prevalent, first as the
pre-relativistic Schroedinger wave function, then as the
Klein-Gordon boson field, providing an early, but later
discarded, model of a ``meson''. Of course, Dirac's spinor took
over as the basic field for ``permanent'' particles as fermions,
with force-field carriers  such as photons, being
bosons. Of  course, the photon field is a vector, not a scalar.
However, investigations of internal spaces for particle
symmetries directly involve gauge theories of force fields. In
this model the internal symmetry spaces for families
of fields have interesting transformation
properties from the internal gauge group viewpoint but are
nonetheless spacetime scalars. Some of the earliest  are  the
$SO(N)$ bosons of the dual model, the Nambu-Goldstone bosons and
the famous Higgs fields.
Of course, the motivations for considering quantum scalar fields
is  certainly very different from those leading to
the scalar field in scalar-tensor theories. Nevertheless, certain
forms of the quantum formalism, and perhaps its macroscopic
manifestations may turn out to be not too different from the
classical scalar fields. Such comparisons may be most evident
in the context of cosmological quantum particle models,
\cite{Kai:05}. We begin with the quantum origin of ``dilatons.''
\par The late 1960's and early 1970's saw the birth of
  quantum dual models, which eventually led to string theory and
later superstring theory, \cite{gsw}. These theoretical models
quite naturally lead to  a scalar field referred to as a
``dilaton.'' This field couples directly  to the
trace of the two-dimensional string stress tensor. This coupling  breaks the Weyl conformal  symmetry of
the string. Since conformal metric transformations
are dilations, we arrive at the word ``dilaton.'' The dilaton
turns out to be what is needed to balance the quantum anomalies
of this tensor by way of beta functionals of this tensor.  In
this analysis,  the Einstein equations for the enveloping
spacetime metric are  ``derived'' as the beta functions.  This
rather involved arguments is discussed in the first volume of the
book  \cite{gsw} which thus provides useful
description of the origin
and role of dilatons. Here we only briefly  summarize the argument in the following. \par
Start with  a  string action as a
natural generalization of a point particle action. Given a
background metric, $g_{\alpha\beta},$ an obvious  choice
is
\begin{equation}
S_1={-1\over 4\pi\alpha'}\int d^2\sigma \sqrt{\vert
h\vert}h^{ab}\partial_aX^\alpha\partial_bX^\beta
g_{\alpha\beta}(X^c),\label{str1}\end{equation}
  with internal coordinate area
$d^2\sigma,$
internal string metric, $h_{ab},\ a,b...=1,2,$
and $\alpha'$  a tension related coupling parameter.
Comparing
$S_1$ to  a relativistic point particle action, we see the need
for an intrinsic surface metric, $h_{ab}$ for the string that is
not present for point particle. Now, assume that the derived
physics should  be independent of the internal
parameterization, that is the choice of string
metric. However,
 any two-dimensional
metric is  conformally flat (but only locally, in general!),
 \begin{equation}
h_{ab}=\phi\eta_{ab},\label{str3}\end{equation}
 with constant $\eta_{ab}.$  So
the surface element appearing in (\ref{str1}) reduces to    the
flat one,
 \begin{equation}
  d^2\sigma\sqrt{\vert
h\vert}h^{ab}=d^2\sigma\eta^{ab}.\label{str4}\end{equation}
\par
In addition to $S_1$, other terms have been proposed. One of
these makes use  of the string geometry through its
curvature scalar,
\begin{equation}
\chi={1\over 4\pi}\int d^2\sigma \sqrt{\vert
h\vert}R^{(2)}.\label{str4.x}\end{equation}
Of course, one of the earliest discoveries
relating geometry and topology was that this integral depends
{\em only} on the topology of the string surface, and not the
particular geometry.  In fact, (\ref{str4.x}) defines the first
Chern class for two dimensions.  The value for $\chi$ is the
Euler number of the surface, and cannot be a dynamical variable.
However, dynamics can be restored by modifying the form of
(\ref{str4.x}) by adding to (\ref{str4.x})  a scalar
field factor, the ``{\bf dilaton},'' $\Phi$, giving
\begin{equation}
S_2={1\over 4\pi}\int d^2\sigma\sqrt{\vert
h\vert}\Phi(X^c)R^{(2)}.\label{str5}\end{equation}
Classically this term
 breaks  the conformal invariance.   However,
perhaps surprisingly, it is precisely this term which can restore
conformal invariance after quantization. When the action
$S=S_1+S_2$ is quantized, conformal invariance is broken (an
anomaly) unless the external fields satisfy three equations. This
argument is described in detail in GSW, volume 1, page 180. Here
we drop the  $B_{\alpha\beta}$ for simplicity and
get (in the magical string dimension 26!)  Einstein-like
equations,
\begin{equation}
0=R_{\alpha\beta}-2\Phi_{;\alpha;\beta},\label{str6}\end{equation}
\begin{equation} 0=4\Phi_{,\alpha}\Phi^{,\alpha}-
4\Phi^{;\alpha}_{\ ;\alpha}+R.\label{str7}\end{equation}
This ``derivation'' of the Einstein equations from string theory
was one of the attractive features of string theory. Recall,
however, that this required the introduction of a dilaton,
spacetime scalar, field to break conformal invariance, which is
later restored only if Einstein-like equations are satisfied.
\par
Now, without regard for their string theory origins,
field equations can be derived
from an ``effective
action,'' \begin{equation}
\delta\int
d^DXe^{-2\Phi}(R-4\Phi_{,\alpha}\Phi^{,\alpha})=0.\label{str8}\end{equation}
Of course, this action is nothing but  a special case
of the vacuum scalar-tensor  one, (\ref{bd-1}),  with
$-2\Phi=\ln\phi,$ and $\omega=1.$  While the
motivation and physics of the scalar field in the
classical, pre-quantum, scalar-tensor theories is
vastly different from the dilaton scalar field, it
is difficult not to notice the close parallel
between the universally coupled scalar of the old scalar-tensor
theories
and the new dilaton. \par
 \section{Inflatons}
 We will not attempt to review the rapidly expanding field of
 rapidly expanding (accelerating) cosmological models, but
end this paper with a few comments about the early days of
inflationary cosmology.\par
Standard
general relativity has long been known to have difficulties in
its application to observed cosmological facts.  For
example, standard general relativity requires that the initial
big bang conditions  be fantastically fine-tuned in order to
result in the universe as we now see it some $10^{11}$ years
later.  See for example Peebles \cite{PJEP}, Linde \cite{al}.
Look at the standard Robertson-Walker isotropic homogeneous
metric model,
\begin{equation}
ds^2=-dt^2+R(t)^2d\sigma_\epsilon^2,\label{cosm1}\end{equation}
where the three-space metric, $d\sigma_\epsilon^2$, is
hyperbolic, flat, or spherical depending on whether $\epsilon$
is -1, 0 or +1. The Einstein equations result in
\begin{equation}
\big({\dot R\over R}\big)^2=\kappa\rho/3+ {\epsilon\over
R(t)^2}+\Lambda/3.\label{dc1}\end{equation}
Defining the Hubble variable as usual, this can be rewritten,
\begin{equation}
1= \Omega+\epsilon\Omega_R+\Omega_\Lambda,\label{cosm2}\end{equation}
where
 \begin{equation}
 \Omega\equiv {\kappa\rho\over
3H^2},\label{dc2}\end{equation}
 \begin{equation}
  \Omega_R\equiv {1\over (R
H)^2},\label{dc3}\end{equation}
and
 \begin{equation} \Omega_\Lambda\equiv
{\Lambda\over 3H^2}.\label{dc4}\end{equation}As of the 1980's   these three quantities were measured to be
each in the ballpark of one. In fact, \begin{equation}
\Omega(now)\approx {\kappa M\over R}\approx 10^0,\label{cosm4}\end{equation}
which is one of Dirac's large number coincidences which was
so instrumental in leading to the scalar-tensor theories.
However, if we stick to standard GR, not a scalar-tensor
variation, $\kappa$ is constant, (\ref{cosm4}) is valid only
now, and takes this value now only if the universe evolves from
very finely tuned earlier values. For example, in the present
matter dominated era the equation of state leads to
\begin{equation}
\rho R^3=M\approx const,\label{cosm7}\end{equation}
whereas in an earlier radiation dominated state
\begin{equation}
\rho R^4\approx const.\label{cosm8}\end{equation}
An analysis of the time evolution of these quantities in standard
general relativity under drastically different regimes show that
an extremely small variation of the values of the $\Omega$'s at
early times would result in drastically different values now.
But this is not the only conceptual problem.  For example, there
are questions of how the universe could have homogenized itself
from random early data (the ``horizon'' problem), and others,
\cite{PJEP},\cite{al}\par
Guth\cite{G} pointed out that this myriad of difficulties could
be at least partially resolved if the early stages of evolution
were ``inflationary,'' that is
\begin{equation}
R(t)=R(0)e^{Ht},\label{cosm9}\end{equation}
with constant $H.$  Such a model is consistent with (\ref{dc1})
for $\rho=\epsilon=0,\ \Lambda\ne 0.$  Of course, this is not
consistent with present data, so something other than a
cosmological constant is needed. One way to achieve it is to
introduce a new massless scalar field, the
``inflaton,'' $\phi,$ with Lagrangian density, \begin{equation}
{\cal L}=g^{\alpha\beta}\phi_\alpha\phi_\beta-V(\phi).\label{inf2}\end{equation}
This field contributes an effective mass density
and pressure given by \begin{equation} \rho_\phi=\dot\phi^2/2+V,\
\ p_\phi=\dot\phi^2/2-V.\label{inf3}\end{equation}
The introduction of $\phi$ and its  potential, $V$, can be used
to resolve at least some, but certainly not all, of the problems
discussed above.  In some models, this inflaton  has a
 dilaton-like nature, in others it is reminiscent of the
$\phi$ in the old scalar-tensor theories. The problems
of scalar-tensor field theories with
solar-system sized observations require that  $\omega$ be very
large. However, this restriction need not diminish the
significance of the inflaton field in earlier cosmological
contexts.  \par
Of course, as of the beginning of the 21st century,
cosmological observations and theory have expanded well past
these early inflationary models, but we will stop here, and
remind the reader that universally coupled, thus gravitational,
scalar fields are still active players in contemporary
theoretical physics. So perhaps we can say that the  scalar
field is still alive and active, if not always well, in current
gravity research.
\par
  
\end{document}